\documentclass[12pt]{article}

\pdfoutput=1
\usepackage{color}
\usepackage{epsfig, palatino}
\usepackage{pstricks,pst-node,pst-tree}
\usepackage{epic}
\usepackage{mathrsfs}
\usepackage{ae} 
\usepackage[T1]{fontenc}
\usepackage[ansinew]{inputenc}
\usepackage{amsmath}
\usepackage{amssymb}
\usepackage{graphicx}
\usepackage{ulem}
\usepackage{color}
\definecolor{darkblue}{cmyk}{0.9,0.9,0,0}
\usepackage[colorlinks=true,linkcolor=darkblue,citecolor=darkblue,urlcolor=darkblue]{hyperref}
\usepackage{cite}
\usepackage{hyperref}
\usepackage{wasysym}
\usepackage{varioref}
\usepackage{makeidx}
\usepackage[english]{babel}
\usepackage{simplewick}
\usepackage{array}

\newcommand{\comment}[1]{}

\newcommand{\beq}{\begin{equation}}
\newcommand{\eeq}{\end{equation}}
\newcommand{\beqq}{\begin{equation*}}
\newcommand{\eeqq}{\end{equation*}}
\newcommand\beqa{\begin{eqnarray}}
\newcommand\eeqa{\end{eqnarray}}
\newcommand\beqaa{\begin{eqnarray*}}
\newcommand\eeqaa{\end{eqnarray*}}
\newcommand\bea{\begin{array}}
\newcommand\eea{\end{array}}

\newcommand{\neqa}{\nonumber\end{eqnarray}}

\renewcommand{\d}{\partial}

\newcommand{\<}{{\langle}}
\renewcommand{\>}{{\rangle}}

\newcommand{\re}{\relax{\rm I\kern-.18em R}}

\renewcommand{\sp}{p\hspace{-.40em}/}

\definecolor{darkgreen}{rgb}{0.0, 0.45, 0.0}

\def\XXint#1#2#3{{\setbox0=\hbox{$#1{#2#3}{\int}$}
\vcenter{\hbox{$#2#3$}}\kern-.5\wd0}}

\def\su2{{SU(2)}}

\def\[{\left[}
\def\]{\right]}

\def\Li{{\rm Li}_2}

\def\({\left(}
\def\){\right)}
\def\[{\left[}
\def\]{\right]}

\def\<{\langle}
\def\>{\rangle}

\def\i2{\frac{i}{2}}

\def\O{{\mathcal O}}

\def\spi{\relax{\rm \pi\kern-0.5em /}}
\def\sA{\relax{\rm A\kern-0.5em /}}
\def\sp{\relax{\rm p\kern-0.5em /}}
\def\sd{\relax{\rm \d\kern-0.5em /}}
\def\sk{\relax{\rm k\kern-0.5em /}}
\def\sn{\relax{\rm n\kern-0.5em /}}
\def\sl{\relax{\rm l\kern-0.5em /}}
\def\sP{\relax{\rm P\kern-0.7em /}}
\def\sBethe{\relax{\rm \Bethe\kern-0.5em /}}
\def\cN{{\cal N}}

\def\cN{{\cal N}}

\def\cW{{\cal W}}

\def\2F1{\,_2{\rm F}_1}

\makeindex

\begin{document}

\thispagestyle{empty}

\renewcommand{\thefootnote}{\fnsymbol{footnote}}
\setcounter{page}{1}
\setcounter{footnote}{0}
\setcounter{figure}{0}

\begin{center}

$$$$

{\Large\textbf{\mathversion{bold}Resumming the POPE at One Loop
}}

\vspace{1.0cm}

\textrm{Ho Tat Lam$^{{\footnotesize\pentagon}}$ and Matt von Hippel$^{{\footnotesize\pentagon}}$}
\\ \vspace{1.2cm}
\footnotesize{\textit{
$^{\pentagon}$Perimeter Institute for Theoretical Physics,
Waterloo, Ontario N2L 2Y5, Canada\\
}  
\vspace{4mm}
}

\par\vspace{1.5cm}

\textbf{Abstract}\vspace{2mm}
\end{center}

The Pentagon Operator Product Expansion represents polygonal Wilson loops in planar $\cN=4$ super Yang-Mills in terms of a series of flux tube excitations for finite coupling. We demonstrate how to re-sum this series at the one loop level for the hexagonal Wilson loop dual to the six-point MHV amplitude. By summing over a series of effective excitations we find expressions which integrate to logarithms and polylogarithms, reproducing the known one-loop result.
\noindent

\setcounter{page}{1}
\renewcommand{\thefootnote}{\arabic{footnote}}
\setcounter{footnote}{0}

 \def\nref#1{{(\ref{#1})}}

\newpage

\tableofcontents

\parskip 5pt plus 1pt   \jot = 1.5ex
\newpage
\section{Introduction}

The Pentagon Operator Product Expansion (or POPE) has shown itself to be a powerful tool for the calculation of polygonal Wilson loops and their dual amplitudes in planar $\cN=4$ super Yang-Mills~\cite{Basso2013vsa, Basso2013aha,POPE}. Making use of integrability, the POPE computes Wilson loops at finite coupling, presented as an expansion in flux tube states propagating across the loop. Kinematically, this expansion corresponds to expanding around a particular collinear limit.

For quite some time, it was unclear if this expansion could be re-summed to obtain the full kinematic dependence of the amplitude. A partial resummation was achieved in~\cite{Drummond:2015jea}, but it was only with recent work by Luc\'ia C\'ordova that such a resummation was shown to be possible in the limit of weak coupling for all flux tube states~\cite{Cordova:2016woh}. C\'ordova shows that, for the NMHV six-particle amplitude at tree level, it is possible to package all combinations of states that can contribute into single effective excitations, creating a series which can be re-summed to match the full (tree-level) amplitude.

In this work, we extend C\'ordova's calculation to one loop for the MHV case. While the expressions that appear are of comparable complexity, computing a one-loop amplitude in this way allows us to observe the appearance of transcendental functions from the POPE, in a way that should generalize to higher loop orders.

We begin in section \ref{sec:states} by describing the effective one-particle excitations needed for MHV. In section \ref{sec:resum} we show how to re-sum them into the polylogarithmic functions of the one-loop hexagon Wilson loop. Finally, we conclude by discussing how this procedure might be extended to higher loops.

\section{Effective One-Particle States for MHV}
\label{sec:states}

We can start by considering the expression for the hexagon Wilson loop given by the POPE program~\cite{POPE}, as a sum over all possible flux tube excitations:

\beq
\cW_6=\sum_m \frac{1}{S_m}\int\frac{du_1\ldots du_m}{(2\pi)^m}\Pi_{\rm dyn}\Pi_{\rm mat}\,.
\eeq
Here $S_m$ are symmetry factors, $u_i$ are rapidities, and $\Pi_{\rm dyn}$ and $\Pi_{\rm mat}$ are referred to as the dynamical part and matrix part respectively. The dynamical part contains all of this expression's dependence on the coupling, while the matrix part takes care of R-symmetry.

The excitations summed over here are combinations of fundamental excitations: gluons (and gluon bound states), fermions, and scalars. While gluons and scalars can be straightforwardly integrated in rapidity, fermions must be integrated over two different Riemann sheets. On one of these sheets the fermion momentum is large with respect to the coupling, while on the other it is small. Hence we follow prior convention and divide fermion integrations into ``large'' and ``small'' fermions, which can be treated separately.

\begin{figure}
\centering
\includegraphics{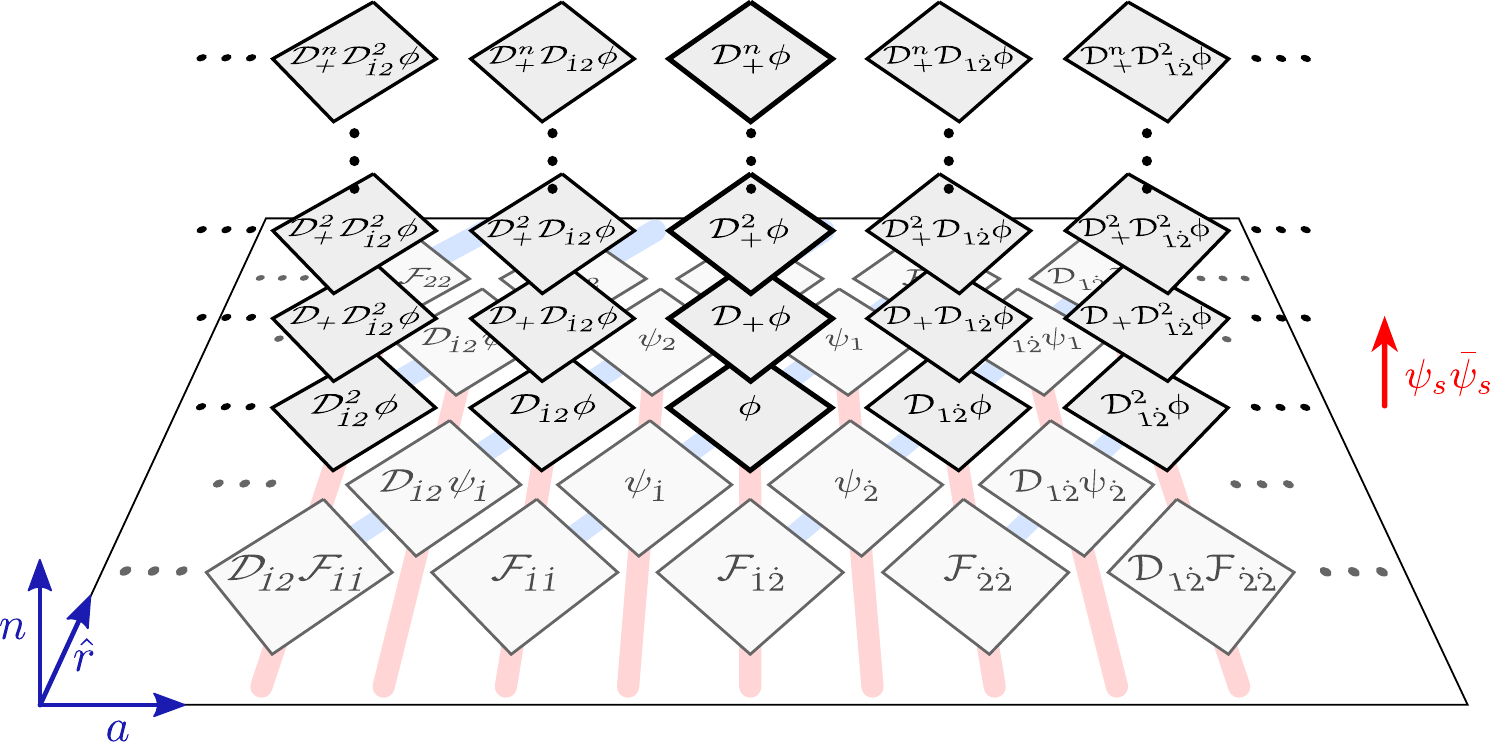}
\caption{Table of effective weak coupling excitations including the first $n$ descendants of the particles transforming in the vector representation of SU(4), from~\cite{Cordova:2016woh}. The plane in the bottom contains the primary excitations.}
\label{descendants}
\end{figure}

Through one loop, only states with one fundamental excitation can contribute, with the exception of small fermions. In practice, then, we can sum effective excitations consisting of one fundamental excitation and a string of small fermions. The small fermion contour allows us to evaluate all small fermion rapidity integrations via residues, so the only integration we need to do explicitly is that of the fundamental excitation. The resulting effective excitations are summarized in figure \ref{descendants}.

While~\cite{Cordova:2016woh} had to consider general R-symmetry representations, here for the MHV case we need only consider the singlets. These correspond to $r=0$ and $r=4$ in the notation of that paper. In particular, we do not need to re-derive the list of residues that must be taken in small fermion rapidity, as Figure 6 of that paper provides the needed information. Specifically, it instructs us to consider ten chains of fundamental excitations and corresponding descendents: $F_b(\psi_S\bar{\psi}_S)^n$, $\psi\bar{\psi}_S(\psi_S\bar{\psi}_S)^n$, $\phi\bar{\psi}^2_S(\psi_S\bar{\psi}_S)^n$, $\bar{\psi}\bar{\psi}^3_S(\psi_S\bar{\psi}_S)^n$, $F_{-b}\bar{\psi}^4_S(\psi_S\bar{\psi}_S)^n$, and their conjugates.

The most straightforward procedure would then be to start with the effective measures given in appendix B of~\cite{Cordova:2016woh}, and find expressions that can be summed over helicity. Instead, we will take a shortcut, and begin with equation (17) of that paper for which this has already been done. For singlet excitations, we specialize to the case where $r_1=4$ and $r_2=0$. There are two cases: the positive helicity excitations (here, just $F_a$) and the rest, which we will refer to as ``gluonic'' and ``non-gluonic''.
\beq
\mu^{[4,0]}_{a,n}(u)|_{\rm gluonic}=\frac{(-1)^{a+n}\Gamma\left(\frac{|a|}{2}-iu\right)\!\Gamma\left(\frac{|a|}{2}+iu\right)}{\Gamma(n+1)\Gamma(|a|+n)}\!\!\left(iu+\frac{|a|}{2}+2\right)_n\!\!\left(iu+\frac{|a|}{2}\right)_n\!\!+\O(g^2)
\eeq
\beq
\mu^{[4,0]}_{a,n}(u)|_{\rm non-gluonic}=\frac{(-1)^{a+n}\Gamma\left(\frac{|a|}{2}-iu-1\right)\!\Gamma\left(\frac{|a|}{2}+iu+3\right)}{\Gamma(n+1)\Gamma(|a|+n+2)}\!\!\left(iu+\frac{|a|}{2}+3\right)_n\!\!\left(iu+\frac{|a|}{2}+1\right)_n\!\!\!+\O(g^2)
\label{eq:nmhvints}
\eeq

For MHV, we have the same excitations, evaluated at the same residues with the same symmetry factors. The only change is in the contributions referred to as NMHV form factors, presented in~\cite{FF}. These are factors present in the NMHV amplitude that take into account the nontrivial R-symmetry of the external states. Since we are interested in the MHV amplitude here, we need to divide the integrands in eq. \ref{eq:nmhvints} by these form factors in order to obtain our desired result.

For the gluonic case, form factors contribute a factor of $h_{F_a}^{-4}(u)$ from the fundamental excitation and a product of contributions from the descendants, where $h_{F_a}(u)$ is the gluonic form factor. Expanded in $g$, $h_{F_a}^{-4}(u)$ is
\beq
h_{F_a}^{-4}(u)=\frac{1}{g^2}\left(\frac{a}{2}+iu\right)\left(\frac{a}{2}-iu\right)+\O(g^0).
\label{eq:glufactors}
\eeq
This tells us two things. First, since the form factor for gluonic excitations starts at order $g^{-2}$, removing it means that what was previously a tree-level NMHV expression now gives us the one-loop expression for MHV. Second, we must also remove a factor of $\left(\frac{a}{2}+iu\right)\left(\frac{a}{2}-iu\right)$.

The contribution from the descendants is also simple to take into account. Expanding the small fermion form-factors $h_{\psi_S}(v)$ in $g$ we find
\beq
h_{\psi_S}^{-4}(v)=\frac{g}{v}+\O(g^2),\quad h_{\bar{\psi}_S}^{-4}(v)=\frac{v}{g}+\O(g^0)
\eeq
Since the descendants consist of pairs of $\psi_S$ and $\bar{\psi_S}$ evaluated at different residues, the factors of $g$ cancel.

For the gluonic case, the first pair of descendants of $F_a$ contains a $\psi_S$ at $u-i\frac{a}{2}$ and a $\bar{\psi_S}$ at $u-i\frac{a}{2}-2i$. Subsequent descendants are at intervals of $i$. Then to leading order in the coupling, the contribution from the descendant form factors is
\beq
\prod_{k=1}^n \frac{u-i\left(\frac{a}{2}+k+1\right)}{u-i\left(\frac{a}{2}+k-1\right)}
\eeq
for $n$ descendants.

Most of these factors cancel. We are left only with contributions from $k=1$, $k=2$, $k=n-1$ and $k=n$. Together, these give an overall factor of
\beq
\frac{\left(u-i\frac{a}{2}-in\right)\left(u-i\frac{a}{2}-in-i\right)}{\left(u-i\frac{a}{2}\right)\left(u-i\frac{a}{2}-i\right)},
\label{eq:descfactors}
\eeq
which we must remove.

Between eq. \ref{eq:glufactors} and eq. \ref{eq:descfactors} we have all that we need to convert the expressions in \ref{eq:nmhvints} to the corresponding integrands for the MHV case in the gluonic sector. The calculation for the non-gluonic states is similar, and is omitted for brevity. Removing these factors, and simplifying using the definition of the Pochhammer symbol $a_n$, we are left with the following expressions:
\beq
\mu^{\rm MHV, gluonic}_{a,n}(u)=g^2\frac{(-1)^{a+n}\Gamma\left(\frac{|a|}{2}-iu\right)\!\Gamma\left(\frac{|a|}{2}+iu\right)}{\left(\frac{|a|}{2}-iu\right)\!\!\left(\frac{|a|}{2}+iu\right)\Gamma(n+1)\Gamma(|a|+n)}\!\!\left(iu+\frac{|a|}{2}\right)^2_n\!\!+\O(g^4)
\eeq
\beq
\mu^{\rm MHV, non-gluonic}_{a,n}(u)=g^2\frac{(-1)^{a+n}\Gamma\left(\frac{|a|}{2}-iu+1\right)\!\Gamma\left(\frac{|a|}{2}+iu+1\right)}{\left(\frac{|a|}{2}-iu\right)\!\!\left(\frac{|a|}{2}+iu\right)\Gamma(n+1)\Gamma(|a|+n+2)}\!\!\left(iu+\frac{|a|}{2}+1\right)^2_n\!\!+\O(g^4)
\label{eq:mhvints}
\eeq

To sum up, one loop MHV, $\mathcal{W}^{\text{MHV}}=1+g^2\mathcal{W}^{\text{MHV,(1)}}+\mathcal{O}(g^4)$, is given by the following POPE series,
\beq
\begin{aligned}
\cW^{\text{MHV}}=1&+2\sum_{n=0}^{\infty}\int\frac{\mathrm{d}u}{2\pi} e^{-(2n+2)\tau+2iu\sigma} \mu^{\rm MHV, non-gluonic}_{0,n}(u)\\
&+\sum_{a\neq0}\sum_{n=0}^{\infty}\int\frac{\mathrm{d}u}{2\pi} e^{-(|a|+2n)\tau+ia\phi+2iu\sigma} \left[\mu^{\rm MHV, gluonic}_{a,n}(u)+e^{-2\tau}\mu^{\rm MHV, non-gluonic}_{a,n}(u)\right]\,.
\end{aligned}
\eeq

\section{One-loop Resummation}
\label{sec:resum}
The resummation of the MHV POPE series can be carried out following a similar strategy to that employed in~\cite{Cordova:2016woh}. The resummation is performed beginning with an expansion in the collinear limit, $\exp(\tau)\rightarrow\infty$, which is then analytically continued to arbitrary kinematics. The key step is to replace the summation over $n$ with integrations over $t$ using the series and integral representations of hypergeometric functions,
\begin{equation}
\begin{aligned}
_2F_1(a,b,c;z)=\sum_{n=0}^{\infty}\frac{\Gamma(a+n)\Gamma(b+n)\Gamma(c)}{\Gamma(a)\Gamma(b)\Gamma(c+n)}\frac{z^n}{n!}=\frac{\Gamma(c)}{\Gamma(b)\Gamma(c-b)}\int_0^1\mathrm{d}t\ t^{b-1}(1-t)^{c-b-1}(1-tz)^{-a}\\
\end{aligned}
\end{equation}
where this integral representation is valid only when $\text{Re}(c)>\text{Re}(b)>0$. After this replacement, the POPE series is converted into the following expression,
\beq
\mathcal{W}^{\text{MHV,(1)}}=\sum_{a\neq 0}\int^1_0\mathrm{d}t\int^{\infty+i\epsilon}_{-\infty+i\epsilon}\frac{\mathrm{d}u}{2\pi}\left(I_{a}^{\rm gluonic}+I_{a}^{\rm non-gluonic}\right)+2\int^1_0\mathrm{d}t\int^{\infty+i\epsilon}_{-\infty+i\epsilon}\frac{\mathrm{d}u}{2\pi}I_{0}^{\rm non-gluonic}
\eeq
with the following integrands,
\beq
\begin{aligned}
I_{a}^{\rm gluonic} &= \frac{(-1)^a t^{\frac{\left| a\right| }{2}+i u-1} (1-t)^{\frac{\left| a\right| }{2}-i u-1}
   \left(e^{-2 \tau } t+1\right)^{-\frac{\left| a\right| }{2}-i u} e^{-\tau  \left| a\right| +i a \phi
   +2 i \sigma  u}}{ u^2+\left| a\right| ^2/4}
\\
I_{a}^{\rm non-gluonic} &= \frac{(-1)^a t^{\frac{|a|}{2} + i u}(1-t)^{\frac{|a|}{2}  - i u} 
   \left(e^{-2 \tau } t+1\right)^{-\frac{\left| a\right| }{2}-i u-1} e^{-\tau  (\left| a\right| +2)+i
   a \phi +2 i \sigma  u}}{u^2+\left| a\right| ^2/4}
\end{aligned}
\eeq

The integration over $u$ can be evaluated by taking residues at $u=\pm i|a|/2$. Only one of the residues will be picked up depending on how the contour is closed. All of the integrands are of the form $\exp\[if(t)u\right]/u^2$ in the limit of large $u$, with
\beqq
f(t)=2 \sigma -\log \left[\frac{(1+e^{-2 \tau } t)(1-t)}{t}\right]
\eeqq
which has a root between 0 and 1 at,
\beqq
t^*=\frac{1}{2} \left[1-e^{2 \sigma +2 \tau }-e^{2 \tau }+\sqrt{\left(1-e^{2 \sigma +2 \tau }-e^{2 \tau }\right)^2+4 e^{2 \tau }}\right]\,.
\eeqq
When $t\in(t^*,1)$, $f(t)>0$, the contour closes in the upper half complex plane and picks up a pole at $u=i|a|/2$ so that the integration at infinity vanishes. On the other hand, when $t\in(0,t^*)$, $f(t)<0$, the contour closes in the lower half-plane and picks up a pole at $u=-i|a|/2$. The prescription is however different when $a=0$. There, the integration contour is shifted upwards on the complex plane as suggested by the POPE proposal (this allows us to reproduce the correct Riemann sheet for the large fermions, which give the $a=0$ contribution). When $t\in(0,t^*)$, a double pole at $u=0$ is selected and when $t\in(t^*,1)$, no pole is selected.

After integrating over $u$, the integration domain of $t$ breaks into two pieces,
\beq
\begin{aligned}
\mathcal{W}^{\text{MHV,(1)}}=&\int^1_{t^*}{\rm d}t\sum_{a\neq0}\frac{(-1)^a e^{ia\phi-|a|\sigma-|a|\tau}}{|a|}(1-t)^{|a|}\(\frac{1}{t}-\frac{1}{t-1}+\frac{1}{t+e^{2\tau}}\)+
\\
&\int_0^{t^*}{\rm d}t \left[\frac{2f(t)}{t+e^{2\tau}}+\sum_{a\neq0}\frac{(-1)^a e^{ia\phi+|a|\sigma-|a|\tau}}{|a|} \left(\frac{e^{2\tau}t}{t+e^{2\tau}}\right)^{|a|}\(\frac{1}{t}-\frac{1}{t-1}+\frac{1}{t+e^{2\tau}}\)\right]\,.
\end{aligned}
\eeq  
We move the summation over helicity inside the integration. This summation converges in the collinear limit, and it has a closed form, found from the following simple relation,
\beq
\sum_{a=1}^\infty \frac{x^a}{a}=-\log(1-x)\,.
\eeq
We are left with the following integrations to be performed,
\beq
\label{MHVintegration}
\begin{aligned}
\mathcal{W}^{\text{MHV,(1)}}=&\int^1_{t^*}{\rm d}t\log\[1+ e^{i\phi-\sigma-\tau}(1-t)\] \(\frac{1}{t-1}-\frac{1}{t}-\frac{1}{t+e^{2\tau}}\)
\\
&+\int_0^{t^*}{\rm d}t \left[\frac{f(t)}{t+e^{2\tau}}+\log\left(1+\frac{e^{i\phi+\sigma+\tau}t}{t+e^{2\tau}}\right)\(\frac{1}{t-1}-\frac{1}{t}-\frac{1}{t+e^{2\tau}}\)\right]+{\rm h.c.}
\end{aligned}
\eeq
We organize the integrands as follows,
\beq
\begin{aligned}
\mathcal{W}^{\text{MHV,(1)}}=
&\int_0^{t^*}{\rm d}t \log\[t+e^{2\tau}+e^{i\phi+\tau+\sigma}t\]\(\frac{1}{t-1}-\frac{1}{t}-\frac{1}{t+e^{2\tau}}\)+\frac{\rm d}{{\rm d}t}\log\[t+e^{2\tau}\]\log\[\frac{e^{2(\tau+\sigma)}t}{1-t}\]
\\
&+\int^1_{t^*}{\rm d}t\log\[1+ e^{i\phi-\sigma-\tau}-e^{i\phi-\sigma-\tau}t\] \(\frac{1}{t-1}-\frac{1}{t}-\frac{1}{t+e^{2\tau}}\)+{\rm h.c.}
\end{aligned}
\eeq
where some of the integrands are combined into total derivatives and the rest are of the following type,
\beqq
\int{\rm d}t \frac{\log(b+at)}{t+c}=\log(b+at)\log\[\frac{a(t+c)}{ac-b}\]+{\rm Li}_2\(\frac{b+at}{b-ac}\)\,.
\eeqq
The integration is made up of two parts: one depending on $t^*$ and the other one with no dependence on $t^*$, which we refer to as the middle term and the boundary term,
\beq
\begin{aligned}
\label{eq:wbefore}
\cW_{boundary}=&\frac{\pi^2}{6}-2\log\[T\]\log\[-\frac{T^2(F^2S+T)(1+F^2ST+T^2)}{F^2S^3}\]+\Li\[-\frac{F^2T}{S}\]
\\
&-\Li\[\frac{F^2S}{F^2S+T}\]-\Li\[\frac{F^2ST}{1+F^2ST+T^2}\]-\Li\[\frac{1}{1+F^2ST+T^2}\]+{\rm h.c.}
\\
\cW_{middle}=&\frac{\pi^2}{6}+\log\[t+\frac{1}{T^2}\]\log\[-\frac{S^2}{T^2(1+tT^2)}\]+ \log\[\frac{v}{w(1+tT^2)}\]\log\[\frac{t-1}{t(1+tT^2)}\]
\\
&+\log\[\frac{v w^*}{T^2}\]\log\[\frac{F^2S}{(S+F^2T)(ST+F^2+F^2T^2)}\]
\\
&-\Li\[v\]-\Li\[-\frac{F^2T}{S}v\]+\Li\[\frac{F^2}{ST+F^2+F^2T^2}v\]
\\
&-\Li\[w\]+\Li\[\frac{F^2S}{F^2S+T}w\]+\Li\[\frac{F^2ST}{1+F^2ST+T^2}w\]+{\rm h.c.}
\end{aligned}
\eeq
where $S=e^\sigma$, $T=e^{-\tau}$, $F=e^{-i\phi/2}$, $v=(F^2+StT+F^2tT^2)/F^2$ and $w=(F^2S+T-tT)/F^2S\,.$ (Note that $v$ and $w$ are not the dual conformal cross-ratios used in sources like~\cite{Dixon2015iva}, which we refer to here as $u_1$, $u_2$, and $u_3$.) Both terms are symmetric under complex conjugation so we can replace some of the terms by their complex conjugates ($w$ to $w^*$ and so on). The new variables $v$ and $w$ satisfy a few relations when $t=t^*$,
\begin{equation}
\frac{v}{w(1+tT^2)}=1,\quad vw^*=v^*w=\frac{ST+F^2+F^2S^2+F^2T^2+F^4ST}{F^2}\,.
\label{vwrels}
\end{equation}

Using these relations the one-loop MHV expression $\mathcal{W}^{\text{MHV,(1)}}=\cW_{boundary}+\cW_{middle}$ can be further simplified to reach the known expression, 
\beq
\begin{aligned}
\label{eq:wafter}
\mathcal{W}^{\text{MHV,(1)}}=&\frac{\pi^2}{6}+\log[S^2]\log[1+T^2]-\log\[u_1\]\log\[u_3\]+\Li\[u_2\]-\Li\[1-u_1\]-\Li\[1-u_3\]\,,
\end{aligned}
\eeq
where
\beq
\begin{aligned}
u_1=&\frac{F^2S^2}{(1+T^2)\(ST+F^2+F^2S^2+F^2T^2+F^4ST\)}\\
u_2=&\frac{T^2}{1+T^2}\\
u_3=&\frac{F^2}{ST+F^2+F^2S^2+F^2T^2+F^4ST}\,.
\end{aligned}
\eeq

A particularly straightforward way to see this simplification is to use symbol methods, as we illustrate in Appendix \ref{symapp}.

\section{Conclusions and Outlook}

Extending the results of~\cite{Cordova:2016woh}, we have demonstrated how to re-sum the Pentagon Operator Product Expansion at one loop to obtain an MHV amplitude. In particular, we have shown how logarithms and polylogarithms emerge in two ways: from the sum over helicity, and via integral representations of hypergeometric functions.

At higher loop orders (and for one loop NMHV) the integrands we found here multiply sums of polygamma functions. Above one loop, we also need to consider multiple effective excitations. Either will make this procedure more complex, but neither should compromise the core of our program. Going forward, it should be possible either to find appropriate choices of integral representations of these functions (similar to that used for the hypergeometric function) or to take their residues in an explicit infinite sum, in either case making the transcendentality properties of the resummation manifest.

Looking farther afield, we anticipate that it may be possible to re-sum the POPE for finite coupling. Doing so will likely involve an as-yet unknown basis of functions. Nevertheless, hints at this stage indicate that this may be more feasible than one would assume. In particular, summing over descendants reduces the complexity of the needed sums over states dramatically, leaving a much simpler sum over effective excitations.
\\ \\ \\
{\bf Acknowledgments}
\\ \\
We owe special thanks to Luc\'ia C\'ordova, who shared an early manuscript of her NMHV resummation which allowed us to attempt this work. We would also like to thank Pedro Vieira and the POPE team at the PSI Winter School for helpful discussions during early stages of this project. This research was supported in part by Perimeter Institute for Theoretical Physics. Research at Perimeter Institute is supported by the Government of Canada through the Department of Innovation, Science and Economic Development Canada and by the Province of Ontario through the Ministry of Research, Innovation and Science.

\appendix

\section{Simplifying \texorpdfstring{$\cW^{\rm MHV}$}{WMHV} Using the Symbol Map}
\label{symapp}

In this appendix we will show the equivalence of our expression for the one-loop MHV Wilson loop in eq. \ref{eq:wbefore} to the known expression in eq. \ref{eq:wafter} using symbols~\cite{Chen,FBThesis,Gonch}. The symbol maps polylogarithmic functions to tensor products of rational functions. For our purposes we need only the action of the symbol map on the dilogarithm and on products of two logarithms:
\beq
\begin{aligned}
\Li[z]\quad \sim \quad &(1-z)\otimes 1/z\\
\log[x]\log[y]\quad \sim \quad &x\otimes y+y\otimes x
\end{aligned}
\eeq
The symbol map is not one-to-one. In particular, constants vanish under the symbol map, so any two functions that differ by a constant are mapped to the same symbol. 
Symbols obey the following relations:
\beqq
\begin{aligned}
\phi_1\otimes\ldots\otimes\phi_i\phi_j\otimes\ldots\otimes\phi_n=\phi_1\otimes\ldots\otimes&\phi_i\otimes\ldots\otimes\phi_n+\phi_1\otimes\ldots\otimes\phi_j\otimes\ldots\otimes\phi_n,\\ \phi_1\otimes\ldots\otimes\phi_i^a\otimes\ldots\otimes\phi_n&=a(\phi_1\otimes\ldots\otimes\phi_i\otimes\ldots\otimes\phi_n)
\end{aligned}
\eeqq
The symbols of the middle and boundary terms can be simplified by using the above relations along with those in eq. \ref{vwrels}. In the end, we collect symbols with the same first entry and get,
\beq
\begin{aligned}
\cW_{boundary}\ \sim \ & {a \otimes \frac{a^2}{F^2ST^3}} + {b \otimes \frac{b^2}{F^2ST^3}} + {c\otimes-\frac{c}{a T^3}} + {d\otimes-\frac{d}{b T^3}} + {F^2\otimes-\frac{ST^6}{b d F^2}}
\\
& + {S^2\otimes\frac{T^{7}}{S}} + {T\otimes\frac{F^{12}S^{14}}{a^3b^3c^3d^3T^{16}}} + {\(1+T^2\)\otimes\frac{F^2S^2T^2}{a b}}
\\
\cW_{middle}\ \sim \ & {a\otimes\frac{F^2ST^3}{a^2}} + {{b\otimes\frac{F^2ST^3}{b^2}}} + {c\otimes-\frac{aT^3}{c}} + {d\otimes-\frac{bT^3}{d}} + {F^2\otimes-\frac{bdF^2}{ST^6}}
\\
& + {S^2\otimes\frac{abS}{F^2T^7}} + {T\otimes\frac{abcdT^{16}}{F^4S^{10}}}+{\frac{ST+F^2(1+S^2+T^2)+F^4ST}{F^2T^2}\otimes\frac{F^4S^2}{abcd}}
\end{aligned}
\eeq
where $a=1+F^2ST+T^2$, $b=ST+F^2+F^2T^2$, $c=F^2S+T$ and $d=S+F^2T$.
Combining them, the symbol of the full expression is,
\beq
\begin{aligned}
\cW^{\rm MHV,(1)}\ \sim\ S^2\otimes\frac{F^2S^2}{cd}+(1+T^2)\otimes\frac{F^2S^2T^2}{ab}+
\frac{ST+F^2(1+S^2+T^2)+F^4ST}{F^2S^2}\otimes\frac{F^4S^2}{abcd}
\end{aligned}
\eeq
which is exactly the expected symbol for one-loop MHV, as can straightforwardly be obtained from the full expression:
\beq
\begin{aligned}
\cW^{\rm MHV,(1)}=&\frac{\pi^2}{6}+\log[S^2]\log[1+T^2]-\log\[u_1\]\log\[u_3\]+\Li\[u_2\]-\Li\[1-u_1\]-\Li\[1-u_3\]\,.
\end{aligned}
\eeq

The symbol maps transcendental constants to zero, so in principle our expression may differ from the known result by a term proportional to $\pi^2$. However, we can check this constant by taking the collinear limit ($T\rightarrow 0$), in which the one-loop hexagon Wilson loop must vanish. We find that our expression satisfies this, and thus is indeed the correct result for one-loop MHV.


\begin{thebibliography}{99}

\bibitem{Basso2013vsa}
B.~Basso, A.~Sever and P.~Vieira,
Phys.\ Rev.\ Lett.\  {\bf 111}, 091602 (2013)
[arXiv:1303.1396 [hep-th]].

\bibitem{Basso2013aha}
B.~Basso, A.~Sever and P.~Vieira,
JHEP {\bf 1401}, 008 (2014)
[arXiv:1306.2058 [hep-th]].

\bibitem{POPE}
B.~Basso, A.~Sever and P.~Vieira,
arXiv:1508.03045 [hep-th].

\bibitem{Drummond:2015jea} 
  J.~M.~Drummond and G.~Papathanasiou,
  JHEP {\bf 1602}, 185 (2016)
  doi:10.1007/JHEP02(2016)185
  [arXiv:1507.08982 [hep-th]].

\bibitem{Papathanasiou:2014yva} 
  G.~Papathanasiou,
  Int.\ J.\ Mod.\ Phys.\ A {\bf 29}, no. 27, 1450154 (2014)
  doi:10.1142/S0217751X14501541
  [arXiv:1406.1123 [hep-th]].

\bibitem{Papathanasiou:2013uoa} 
  G.~Papathanasiou,
  JHEP {\bf 1311}, 150 (2013)
  doi:10.1007/JHEP11(2013)150
  [arXiv:1310.5735 [hep-th]].

\bibitem{Cordova:2016woh} 
  L.~C\'ordova,
  arXiv:1606.00423 [hep-th].
	
\bibitem{FF}
  B.~Basso, J.~Caetano, L.~Cordova, A.~Sever and P.~Vieira,
  ``OPE for all Helicity Amplitudes II. Form Factors and Data analysis,''
  JHEP {\bf 1512} (2015) 088
  doi:10.1007/JHEP12(2015)088
  [arXiv:1508.02987 [hep-th]].
	
%
\bibitem{Dixon2015iva} 
  L.~J.~Dixon, M.~von Hippel and A.~J.~McLeod,
  JHEP {\bf 1601}, 053 (2016)
  doi:10.1007/JHEP01(2016)053
  [arXiv:1509.08127 [hep-th]].
	
\bibitem{Chen}
K.~T.~Chen, 
Bull. Amer. Math. Soc. 83, 831 (1977).

\bibitem{FBThesis}
F.~C.~S.~Brown, 
Annales scientifiques de l'ENS 42, fascicule 3, 371 (2009)
[math/0606419].

\bibitem{Gonch}
A.~B.~Goncharov,
arXiv:0908.2238v3 [math.AG].
	
	
\end{thebibliography}
\end{document}